\documentclass{aa}

\usepackage[utf8]{inputenc}
\usepackage{tabularx}
\usepackage{multirow}
\usepackage{url}
\usepackage{amsmath}	
\usepackage{amssymb}	
\usepackage{graphicx}
\usepackage{color}
\usepackage{natbib}
\usepackage{hyperref}

\usepackage{times,txfonts}
\usepackage{bm}
\usepackage[T1]{fontenc}
\usepackage{ae,aecompl}

\bibpunct{(}{)}{;}{a}{}{,} 
\DeclareUnicodeCharacter{00A0}{ }

\newcommand*\mean[1]{\overline{#1}} 

\newcommand*\tave[1]{\left[ \overline{#1} \right]}
\newcommand*\zave[1]{\left\langle \overline{#1} \right\rangle}

\renewcommand{\vec}[1]{\ensuremath{\bm{\mathit{#1}}}}
\newcommand{\Bbar}{\mean{\vec{B}}} 
\newcommand{\Bbarsq}{\mean{B}^2}
\newcommand{\Bbm}{\vec{B}}
\newcommand{\bbm}{\vec{b}} 
\newcommand{\Bbx}{\mean{B}_x} 
\newcommand{\Bby}{\mean{B}_y} 
\newcommand{\Bbz}{\mean{B}_z} 

\newcommand{\Abm}{\vec{A}}

\newcommand{\Vbar}{\mean{\vec{V}}} 
 
\newcommand{\Vbm}{\vec{V}} 
\newcommand{\vbm}{\vec{v}} 

\newcommand{\Jbar}{\mean{\vec{J}}}

\newcommand{\Jbm}{\vec{J}}
\newcommand{\Jbx}{\mean{J}_x} 
\newcommand{\Jby}{\mean{J}_y} 

\newcommand{\dt}{\frac{\partial}{\partial t}}

\newcommand\emf{\mathcal{E}}
\newcommand\emfb{\vec{\mathcal{E}}}

\usepackage{color}
\newcommand{\red}[1]{\textcolor{red}{#1}}

\newcommand{\argminB}{\mathrm{argmin}}

\makeatletter
\g@addto@macro{\UrlBreaks}{\UrlOrds}
\makeatother

\begin{document}

\title{Exploring helical dynamos with machine learning: Regularized linear regression outperforms ensemble methods}
\titlerunning{Machine learning and dynamo theory}

\author{Farrukh Nauman$^{1}$\footnote{E-mail: naumanf@chalmers.se}, 
	Joonas N\"attil\"a$^{2}$\footnote{E-mail: joonas.nattila@su.se} \\
	$^{1}$Department of Space, Earth and Environment, Chalmers University, SE-41296 Gothenburg, Sweden. \\
	$^{2}$Nordita, KTH Royal Institute of Technology and Stockholm University, Roslagstullsbacken 23, SE-10691 Stockholm, Sweden
}

\author{Farrukh Nauman\inst{1} \and Joonas N\"attil\"a\inst{2}}
\institute{
	Department of Space, Earth and Environment, Chalmers University, SE-41296 Gothenburg, Sweden. \email{naumanf@chalmers.se} \and
    Nordita, KTH Royal Institute of Technology and Stockholm University, Roslagstullsbacken 23, SE-10691 Stockholm, Sweden \email{joonas.nattila@su.se}
}



\label{firstpage}

\abstract{
We use ensemble machine learning algorithms to study the evolution of magnetic fields in magnetohydrodynamic (MHD) turbulence that is helically forced. 
We perform direct numerical simulations of helically forced turbulence using mean field formalism, with electromotive force (EMF) modeled both as a linear and non-linear function of the mean magnetic field and current density.
The form of the EMF is determined using regularized linear regression and random forests.
We also compare various analytical models to the data using Bayesian inference with Markov Chain Monte Carlo (MCMC) sampling.
Our results demonstrate that linear regression is largely successful at predicting the EMF and the use of more sophisticated algorithms (random forests, MCMC) do not lead to significant improvement in the fits. 
We conclude that the data we are looking at is effectively low dimensional and essentially linear.
Finally, to encourage further exploration by the community, we provide all of our simulation data and analysis scripts as open source \textsc{IPython} notebooks.
}


\maketitle

\section{Introduction} \label{sec:intro}
Large scale magnetic fields are observed on the Earth, the Sun, galaxies. Understanding the origin and sustenance of these magnetic fields is the subject of dynamo theory. Magnetic fields observed in the galaxies are thought to have their origin in the early phases of cosmological evolution \citep{subramanian2019}. On relatively smaller scales, magnetic fields likely play a significant role in star \citep{wurster} and planet formation \citep{ercolano}. For the Sun, a long standing question has been to understand the origin of the cyclical behavior of the magnetic North and South poles \citep{brun2017}. 

A number of outstanding questions remain about dynamo theory. Classical models of dynamo theory \citep{krause1967,krause1980} assume that the background flow is stationary and the magnetic field is too weak to feedback on the flow. This reduces the problem of magnetic field generation to a closure problem: how does the fluctuating electromotive force (EMF) depend on the mean field? Previous work on dynamo theory has often modeled the EMF as a \textit{linear} function of the mean field (see \citealt{brandenburg2005}, for an extensive review). 
This approximation is reasonable in the growth (kinematic) phase in a flow-driven magnetohydrodynamic (MHD) system where the initial magnetic energy is orders of magnitude smaller than the kinetic energy of the forced fluid. But the limitations become apparent when the magnetic fields have gained enough energy to modify the background flow through the Lorentz force. 
To model this backreaction of the mean magnetic field on the flow, various proposals have been put forward including models where EMF is a non-linear function of the mean field \citep{jepps1975,pouquet1976,vainshtein1992}. 

Two methods have been widely used in numerical simulations of dynamos to get an explicit form for the electromotive force and both are essentially linear: the test field method \citep{schrinner2007} and linear regression using the least squares method \citep{bran2002}. The kinematic test field method uses a set of orthogonal `test' fields that are pre-determined and can be used to compute the various coefficients in the EMF expansion  corresponding to terms linear in the large scale field and its gradients. As input, the test field requires both the small and large scale velocity field and dynamically solves (on the fly) the test field equations to compute the test field coefficients. This has the advantage of retaining dynamical information \citep{hubbard2009,rheinhardt2012}. Linear regression, in contrast, is mostly used in post-processing to model the ``inverse'' problem of determining the coefficients corresponding to different terms in the linear expansion of the EMF.

Machine learning and deep learning have revolutionized image processing, object detection, natural language processing \citep{lecun}. Recently there has been a surge of interest in using machine and deep learning tools in dynamical systems and computational fluid dynamics \citep{hennigh17, kutz17, brunton_kutz_2019}. Data-driven modeling is not new to fluid dynamics; in meteorology, observational data has been used for decades to guide weather predictions (see \citealt{navon2009}, for a historical overview). Considerable increase in computational power with advances in hardware (GPUs) has allowed modern optimization methods to become increasingly tractable. The end goal is to not only build predictive data-driven models but also to understand underlying complex physics that cannot be captured by simple linear regression.

The motivation to study reduced descriptions of the full 3D MHD simulations comes primarily from the extremely large, and currently numerically intractable, magnetic Reynolds number of most astrophysical systems ($Rm \sim 10^6-10^{12}$, \citealt{brandenburg2005}). Reduced-order modeling of dynamical systems is currently a highly active field of research that aims to provide description of physical systems in terms of the minimal possible degrees of freedom \citep{kutz2013,kutz17}. In this paper, we attempt to build reduced-order models for MHD dynamos by applying non-linear regression tools to the widely studied $\alpha^2$ dynamo \citep{brandenburg2005}. 
In this context, we follow the best practices in machine learning that emphasize generalization by using train-validation-test splits in data. 
If one uses all the data to compute correlations (or use any other algorithm for that matter), it amounts to `descriptive' modeling.
By separating the data into distinct training, validation and test sets, one enters into the realm of `predictive' modeling where the generalization power of different algorithms can be tested by first fitting data on the training set, and then testing it on the validation before making predictions on the test set.
A high score on the training set but a poor score on validation set indicates poor generalization. 
This is in contrast to standard linear regression techniques employed in the dynamo literature where the entire data set is taken to be the training set with no validation set to test overfitting \citep{bran2002}.

We use data from Direct Numerical Simulations (DNS) of forced helical turbulence to construct a reduced model where the EMF is an unknown (linear and non-linear) function of the large scale field. This allows us to quantitatively assess the standard assumption of modeling EMF as linearly proportional to the mean magnetic field and current density. We employ three classes of modern statistical and machine learning models: (i) Linear regression, (ii) Random forests, (iii) Bayesian inference (with Markov Chain Monte Carlo model minimization).
Our selection of a well-studied systems like forced helical MHD turbulence supports our second main motivation: we can compare our results from machine learning tools against relatively well-known results in the literature. The insights gained from such a comparison could then help guide future work on more complex MHD turbulent systems such as MHD turbulence with differential rotation \citep{blackmanbrandenburg2002,charbonneau}.

We describe our data and numerical simulations in the next section. This is followed by a section that begins by explaining the fitting methods considered in this paper. 
We then apply these machine learning and statistical methods on the data. We discuss our results in Section 4 putting out work in the context of previous work and some caveats. Section 5 is conclusions.
All of our analysis scripts and data are freely available as interactive \textsc{IPython} notebooks.%
\footnote{%
    \url{https://github.com/fnauman/ML_alpha2}
}

\section{Data}
We describe the equations, setup and code that was used to conduct DNS of forced turbulence in magnetized flows.

\subsection{Description of DNS}

We use the publicly available 6th order finite difference \textsc{pencil-code} \citep{pencilcode}\footnote{\url{http://github.com/pencil-code}}. 
The MHD equations are solved in a triply periodic cubic domain of size $(2\pi)^3$:
\begin{align}
	\dt \rho &= -\nabla\cdot(\rho \Vbm) \label{eq:density} \\
    \dt \Vbm &= -\frac{1}{\rho} \left[ \Vbm \cdot \nabla \cdot \Vbm + \nabla p - 2\nabla \cdot (\nu\rho\vec{S}) - \Jbm\times\Bbm \right] + \vec{f} \label{eq:NS}\\
	\dt \Abm &= \Vbm\times\Bbm -\mu_0 \eta \Jbm \label{eq:ind_eq},
\end{align}
where $\nu$ is the viscosity, $\vec{S} = \frac{1}{2} (V_{i,j}+V_{j,i}) - \frac{1}{3} \delta_{ij} \nabla\cdot\vec{V}$, $\vec{f}$ is the forcing function, $\eta$ is the resistivity, and $\Jbm = \mu_0^{-1} \nabla\times\Bbm$ is the current density.

The forcing function is taken to be homogeneous and isotropic with explicit control over the fractional helicity determined by the $\sigma$ parameter. It is defined as:
\begin{equation}
    \vec{f}(t,\vec{r}) = Re\left\{ N\vec{f}_{\vec{k}(t)}\exp[i\vec{k}(t)\cdot\vec{x} + i\phi(t)] \right\},
\end{equation}
where the wavevectors $\vec{k}(t)$ and the phase $|\phi(t)|<\pi$ are random at each time step. The wavevectors are chosen within a band to specify a range of forcing centered around $\vec{k}_f$. The normalization factor is defined such that the forcing term matches the physical dimensions of the other terms in the Navier Stokes equation, $N = f_0 c_s(|\vec{k}| c_s/\delta t)^{1/2}$, where $f_0$ is the forcing amplitude, $c_s$ is the sound speed, $\delta t$ is the time step. The Fourier space forcing has the form \citep{haugen2004}:
\begin{equation}
    \vec{f}_{\vec{k}}=\vec{R}\cdot\vec{f}_{\vec{k}}^{\mathrm{nohel}}\quad\text{with}\quad R_{ij}={\delta_{ij}- \frac{i\sigma\epsilon_{ijk}\hat{k}_k} {\sqrt{1+\sigma^2}}},
\label{eq: forcing}
\end{equation}
where $\sigma$ is a measure of the helicity of the forcing; for positive maximum helicity, $\sigma=1$, and 
$\vec{f}_{\vec{k}}^{\mathrm{nohel}} = \vec{k}\times\hat{\vec{e}}/\sqrt{k^2 - (\vec{k}\cdot\hat{\vec{e}})^2}$.

\begin{table}
\begin{center}
\begin{tabular}{ c|c|c|c|c } 
 Name & $c_s/(k_1 \eta)$ & $Rm$ & $t_{res}$ & $v_{rms}$ \\
 \hline
 R5e2 & 500   & 1.68 & 4.97 & 0.0337 \\ 
 R1e3 & 1000  & 4.44 & 9.94 & 0.0446 \\ 
 R2e3 & 2000  & 10.31 & 19.88 & 0.052 \\ 
 R3e3 & 3000  & 16.71 & 30.12 & 0.055 \\ 
 R4e3 & 4000  & 22.64 & 39.76 & 0.057 \\ 
 R5e3 & 5000  & 28.72 & 49.70 & 0.058 \\ 
 R6e3 & 6000  & 34.61 & 59.63 & 0.058 \\ 
 R7e3 & 7000  & 40.41 & 69.58 & 0.058 \\ 
 R8e3 & 8000  & 46.22 & 79.52 & 0.058 \\ 
 R9e3 & 9000  & 50.16 & 89.55 & 0.056 \\ 
 R1e4* & 10000 & 55.79 & 99.40 & 0.056 \\
 R15e4* & 15000 & 82.12 & 149.1 & 0.055 \\
\end{tabular}
\caption{
    Summary of the simulations used in this paper. 
    We used a resolution of $256^3$ and a box size of $(2\pi)^3$ for all of our simulations.
    The forcing scale $k_f/k_1=10$ is used for all runs. 
    Shock viscosity was used in starred (*) simulations.}
\label{table:simulations}
\end{center}
\end{table}

The initial velocity and density are set to zero while the magnetic field is initialized through a vector potential with a small amplitude ($10^{-3}$) Gaussian noise. 
We define the units \citep{brandenburg2001}:
\begin{equation}
c_s = \rho_0 = \mu_0 = 1,
\end{equation}
such that the magnetic field, $\vec{B}$, is in the units of Alf\'ven speed ($c_s \sqrt{\mu_0 \rho_0}$). The domain size is $L^3 = (2\pi)^3$, which leads to $k_1 = 2\pi/L = 1$. Moreover, we use the following dimensionless numbers:

\begin{equation}
Rm = v_{\mathrm{rms}}/k_f\eta, \quad t_{\mathrm{res}} = 1/k_1^2 \eta, 
\end{equation}
where $k_f/k_1 = 10$ is the forcing wavemode.
We provide a summary of the simulations in Table \ref{table:simulations}.

\subsection{Reduced model}
Numerical simulations of large scale astrophysical flows with magnetic Reynolds number well in the excess of $10^6$ are prohibitive. Moreover, state-of-the-art simulations requires $\mathcal{O}$($10^6$) CPU hours to compute for moderate Reynolds numbers. To make modeling such systems more tractable, it is therefore necessary to look for simpler surrogate models that can capture the most significant properties of the high fidelity simulations (DNS). 

\begin{figure*}
	\centering
	\includegraphics[trim={1.3cm 0.0cm 0.4cm 0},clip,width=0.42\textwidth]{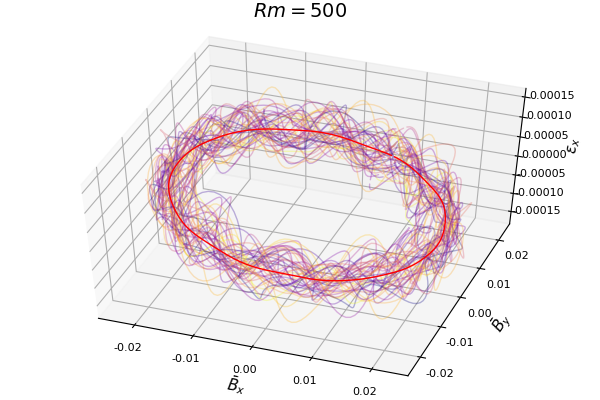}
	\includegraphics[trim={1.3cm 0 0.4cm 0},clip,width=0.42\textwidth]{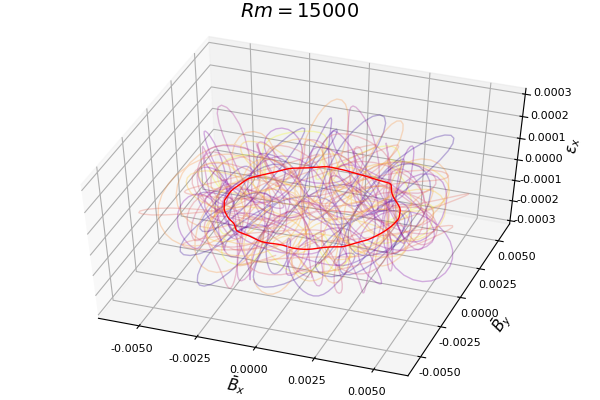}
	\caption{
        3D parametric visualization of the $\Bbx(t,z)$ vs. $\Bby(t,z)$ and $\emf_x(t,z)$ fields for different time slices as a function of $z$ (purple curves). The thick red curves show the time averaged (beyond the kinematic regime) vertical profiles for $Rm = 500$ (left panel), and $1.5 \times 10^4$ (right panel).
		The small $Rm$ simulations have a smoothly evolving circular dependencies while higher $Rm$ cases develop secondary oscillations around the main ``torus''.
		This is consistent with the simulation becoming more ``turbulent'' as Reynolds number is increased.
	}
	\label{fig:3Dfields}
\end{figure*}

Large scale dynamo theory concerns itself with the evolution equation of the filtered induction equation, 
\begin{equation}
	\dt \Bbar = \nabla \times (\Vbar\times\Bbar) + \nabla \times \emfb + \eta \nabla^2 \Bbar,
	\label{eq:ind}
\end{equation}
where EMF is given as $\emfb = \overline{\vbm \times \bbm}$, $\eta$ is the turbulent resistivity, and the overbar represents a filtered quantity. The filter could be ensemble, temporal or spatial depending on the problem. 
The first term on the right-hand side is typically ignored unless there is a strong non-constant mean velocity field such as in the case of galaxies (the mean differentially rotating flow for the galactic dynamo is considered in \citealt{shukurov2006}). 
The dynamics of the velocity field (Eq. \ref{eq:NS}) are ignored based on the assumption that the forcing term is the dominant term and it causes the velocity field to be in a steady state. 
This assumption is only reasonable in the `kinematic' regime where the magnetic fields are too weak to modify the velocity field dynamically.

In keeping with previous work that has used horizontal averaging as the filter in Eq.~\eqref{eq:ind} \citep{brandenburg2001, brandenburg2005}, our data has been averaged over the whole $xy$ plane:
\begin{equation}
	\Bbar = \frac{1}{L_x L_y} \int_{0}^{L_y}\int_{0}^{L_x} \vec{B} dx dy.
\end{equation}
This averaging scheme has the interesting property that $\nabla\times(\Vbar\times\Bbar) \sim 0$ since $\Bbz = 0$ (due to incompressibility and periodic boundary conditions). We only consider the fields (let $Q$ be an arbitrary field) reduced to 1D by furthermore averaging them in two different ways:
\begin{enumerate}
	\item $z$ averaged: $\zave{Q}(t) = 1/(z_2 - z_1) \int_{z_1}^{z_2} \mean{Q} dz$ where $z_1$, $z_2$ are chosen such that they do not cover the entire z-domain since that will correspond to volume averaged quantities that are zero.
	\item $t$ averaged: $\tave{Q}(z) = 1/(t_2 - t_1) \int_{t_1}^{t_2} \mean{Q} dt$ where $t_1$, $t_2$ are chosen to correspond to either the kinematic or saturation regime for different runs.
\end{enumerate}

This leaves two independent magnetic fields and corresponding EMF components. 
Analytical models that use $\emfb = \alpha \Bbar - \red{\eta_t} \Jbar$ are termed $\alpha^2$ dynamos and have been extensively studied in the literature. 
ere $\eta_t$ represents the turbulent resistivity.

In Fig.~\ref{fig:3Dfields}, we show a three-dimensional phase-space visualization of $\emf_x$ vs. $\Bbx$ vs. $\Bby$ that shows the relation between the EMF and the mean magnetic fields (situation is the same when considering $\emf_y$ too). 
The circle in $\Bbx-\Bby$ plane represents the fact that the total magnetic energy $\Bbx^2 + \Bby^2 \sim \text{constant}$. 
The slight tilt with respect to the EMF (i.e., towards $z$ direction) implies a linear correlation with the EMF. 
The increasing disorder with increasing $Rm$ is characteristic of highly turbulent systems.

\section{Model fits} 
We describe the 1D linear and non-linear regression for temporal and vertical profiles in the following subsections. 
In the following, we will refer to the magnetic fields as both ``fields'' and ``features'', and EMF as ``target'' interchangeably. For most of the following, we will focus on $\emf_x$ but similar results hold for $\emf_y$ since we are looking at isotropically forced MHD turbulence with no background field.

\subsection{Description of algorithms considered in this paper}
Ordinary Least Squares (OLS) assumes that the data is linear, features/terms are independent, variance for each point is roughly constant (homoscedasticity), features/terms do not interact with one another. Moreover, causal inference based on linear regression can be misleading \citep{Bollen2013}, a problem shared by all curve fitting methods. We should highlight the distinction between \textit{linearity} of the regression as opposed to \textit{linearity} of the features. For example, one can construct a non-linear basis (example: $x, x^2, x^3, ...$ or $\sin{x}, \cos{x}, ...$) and do linear regression with this basis. Linear regression on non-linear data can potentially represent non-linear data well assuming that the non-linear basis terms are independent and have low variance. Constructing the correct interacting/non-linear terms requires domain expertise. For this reason, it is important to also consider non-linear regression methods such as random forests that are robust to noise and are capable of modeling interactions implicitly.

\subsubsection{LASSO}
Most real data have considerable variance that can lead to misleading fits from linear regression.
Regularized linear regression aims to construct models that are more robust to outliers. Two of the most common regularization schemes for linear regression are \citep{tibshirani1996,hastie2015}: (i) LASSO (Least Absolute Shrinkage and Selection Operator; also known simply as L1 norm), (ii) Ridge (also known as L2 norm). As opposed to ridge, LASSO has the advantage of doing feature selection by reducing coefficients of insignificant features. We will use LASSO regression in this work.

Intuitively LASSO reduces variance at the cost of introducing more bias. 
LASSO works better than ridge regression if the data is low dimensional (or the coefficient matrix is sparse). 
For high dimensional complex data, neither ridge nor LASSO do well. Because of this, the ``Bet on sparsity" principle \citep{hastie2008} suggests trying LASSO for all datasets. 
Formally, LASSO optimizes for:
\begin{equation}
\argminB_w \left[ || y - X w ||_2^2 + \alpha || w ||_1 \right],
\end{equation}
where $y$ represents the target vector (in our case, $\emfb$), $\vec{X}$ is the feature matrix ($B_x, B_y, ...$ are the columns), $w$ is coefficient vector and $\alpha$ is a parameter with a range between $0$ and $1$. 
The subscripts `1' and `2' represent L1 ($\sum_n ||w||_1 = |w_1| + |w_2| + ... + |w_n|$) and L2 ($\sum_n ||w||_2 = \sqrt{w_1^2 + w_2^2 + ... + w_n^2}$) norms, respectively. 
A higher $\alpha$ penalizes outliers more strongly and shrinks coefficients of unimportant features to zero while a lower $\alpha$ is similar to linear regression with no regularization (OLS).

\subsubsection{Random forests}
Random forests belong to the class of ensemble machine learning algorithms \citep{breiman2001,hastie2008}. 
A random forest consists of a large number of decision trees that each take a random subset of features with a bootstrapped sample of the data \citep{raschka2018}. 
This helps in getting rid of strong correlations among different trees but some correlations might still remain. 
Random forests are one of the most popular and widely used machine learning algorithms because they require little hyperparameter tuning and can handle large noisy data sets. 
From a computational point of view, each decision tree in the ensemble is independent and can thus be easily processed in parallel. 
Moreover, because of the weighted average over the ensemble, random forests are robust to strong variance in the data but a few deep decision trees can lead to strong bias. 

Ensembles of decision trees (random forests and gradient boosting) are non-parametric models meaning they do not require information about the underlying data distribution. In other words, unlike in linear regression where the shape of the function that connects the independent variables to the dependent variable is fixed, the functional form is not known in random forests and is instead determined through training. This offers a different perspective on modeling, termed `algorithmic' modeling as opposed to `data' modeling \citep{breiman2001two}. The non-parametric nature of random forests and gradient boosting make them applicable to a wide class of problems since in most real-world situations, data distribution is not known a-priori. However, this comes at the cost of interpretability making ensembles of decision trees harder to understand than linear regression \citep{strobl2008}.

Feature importance and selection: As opposed to linear regression, random forests do not directly measure the coefficient of the features. 
A single decision tree is highly interpretable but a whole ensemble of decision trees is harder to interpret \citep{breiman2001two}. 
Random forests offer some insight through feature importances \citep{hastie2008}. We use the ``mean decrease impurity" method implemented in \textsc{Python} library \textsc{scikit-learn}  \citep{scikit-learn} \citep{breiman1984,louppe2014}. With this method, the feature importance corresponds to the frequency of a particular feature appearing across different trees combined with the reduction of error corresponding to that particular feature, weighted by the fraction of samples in that particular node. One popular alternative is to randomly permute the data for a particular feature and see whether it has an effect on the mean squared error (or information gain/gini importance for classification tasks). If the mean squared error increases or remains the same, it implies that the particular feature is not important. This is termed as ``permutation'' importance \citep{breiman2001, strobl2008}. 

\subsubsection{Bayesian inference}

Bayesian inference is a powerful statistical framework that presents an alternative way to perform the model parameter estimation.
The main strengths of Bayesian-based approaches are their ability to quantify the uncertainty of the model and the possibility to incorporate previous a-priori knowledge into the estimation.
The downside, when compared to many conventional machine learning methods, is that the model needs to be known before the fit.

Let us quickly summarize the idea behind Bayesian inference.
In general, let us present our model with $\mathcal{M}$ and our empirical (simulation) data with $\mathcal{D}$.
We are then interested in how well can our model describe the data, or quantitatively what is the probability that our model is valid given the data, $\Pr(\mathcal{M}|\mathcal{D})$.
By directly comparing our model and the data, what we actually get is, however, the probability of our model given the data, known as the likelihood $\mathcal{L} = \Pr(\mathcal{D}|\mathcal{M})$.
Our original question of the validity of our model, can be answered by applying the Bayes' theorem presented as (see e.g., \citealt{grinstead97})
\begin{equation}\label{eq:bayes}
\Pr(\mathcal{M}|\mathcal{D}) = \frac{\Pr(\mathcal{D}|\mathcal{M})\Pr(\mathcal{M})}{\Pr(\mathcal{D})},
\end{equation}
where $\Pr(\mathcal{M})$ is our prior probability of the model and $\Pr(\mathcal{D}) = \int \Pr(\mathcal{D}|\mathcal{M}) d\mathcal{M}$ is the prior probability.
Therefore, because $\mathcal{L}$ can be computed and $\Pr(\mathcal{M})$ is known a-priori, we can use the Bayes' theorem to quantitatively asses the validity of, for example, our $\alpha^2$ dynamo models given the simulation data results.

In practice, not only are we interested in the validity of the model, but also what are the parameters $\Theta$ of the model, $\mathcal{M}(\Theta)$, that best describe the data. 
In Bayesian inference, these model parameters are determined using a marginal estimation where the resulting posterior parameter for model parameter $\Theta_j$ are obtained by integrating over the probability of all the other model parameters, $\Theta_i$ ($i \ne j$).
Then, this (one-dimensional) distribution represents the probability that the $j$th model parameter, $\Theta_j$, will take a particular value given our data $\mathcal{D}$.

Here we solve the Eq.~\eqref{eq:bayes} (and therefore obtain also posterior distributions for our model parameters) using Monte Carlo Markov Chain (MCMC) integration techniques.
To perform the MCMC fit we use the publicly available \textsc{emcee} library \citep{emcee13}.
We employ the affine-invariant stretch-move ensemble sampler with typically $3 N_{\mathrm{param}}$ number of walkers (members of ensemble), where $N_{\mathrm{param}}$ is the number of fit parameters.
No prior knowledge is incorporated into the fits and so we use uniform priors for all of our model parameters.

\subsection{Vertical profiles}
In this subsection, we apply various methods on time averaged data that only has vertical dependence ($256$ grid points). 

\begin{figure*}
	\centering
	\includegraphics[width=0.45\textwidth]{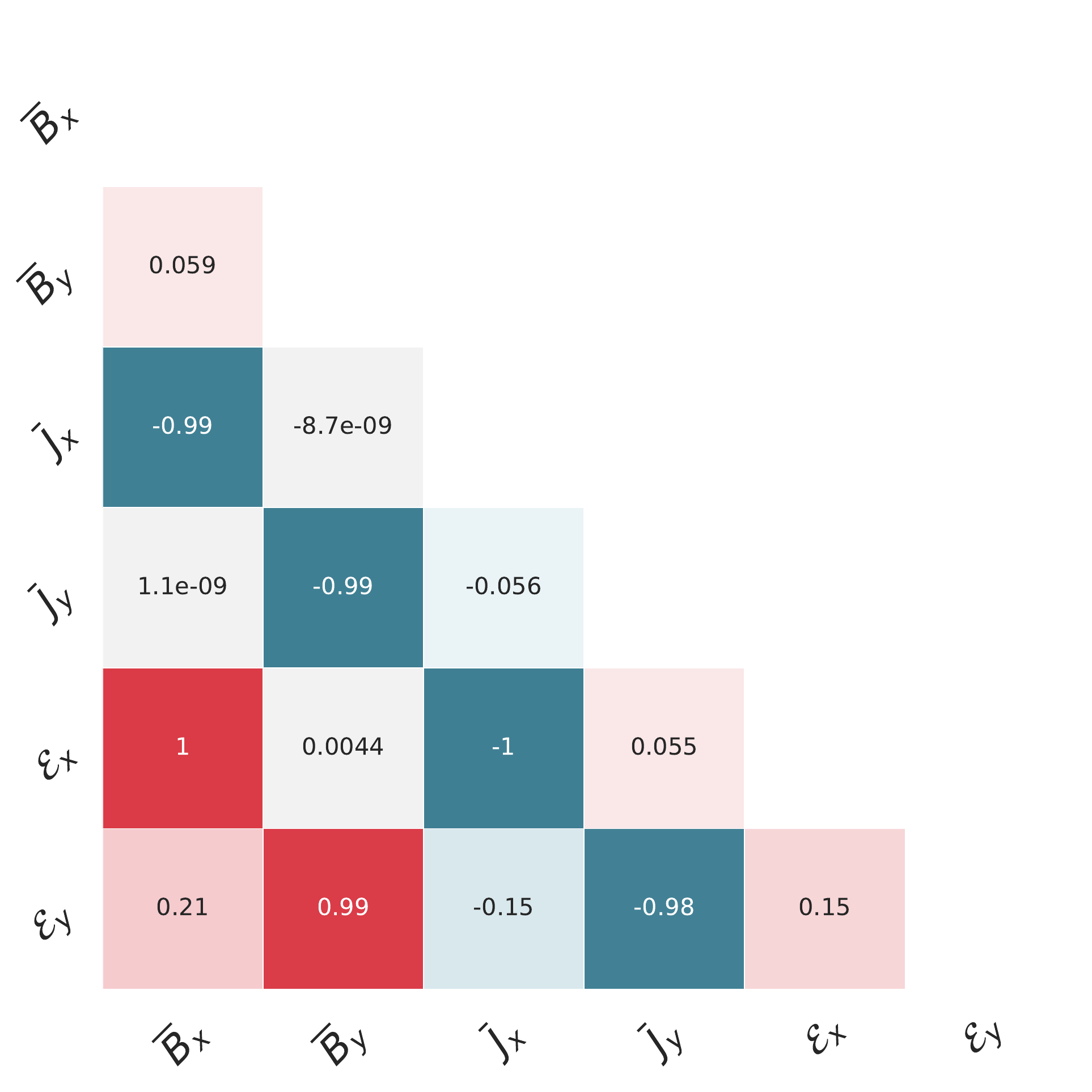}
	\includegraphics[width=0.45\textwidth]{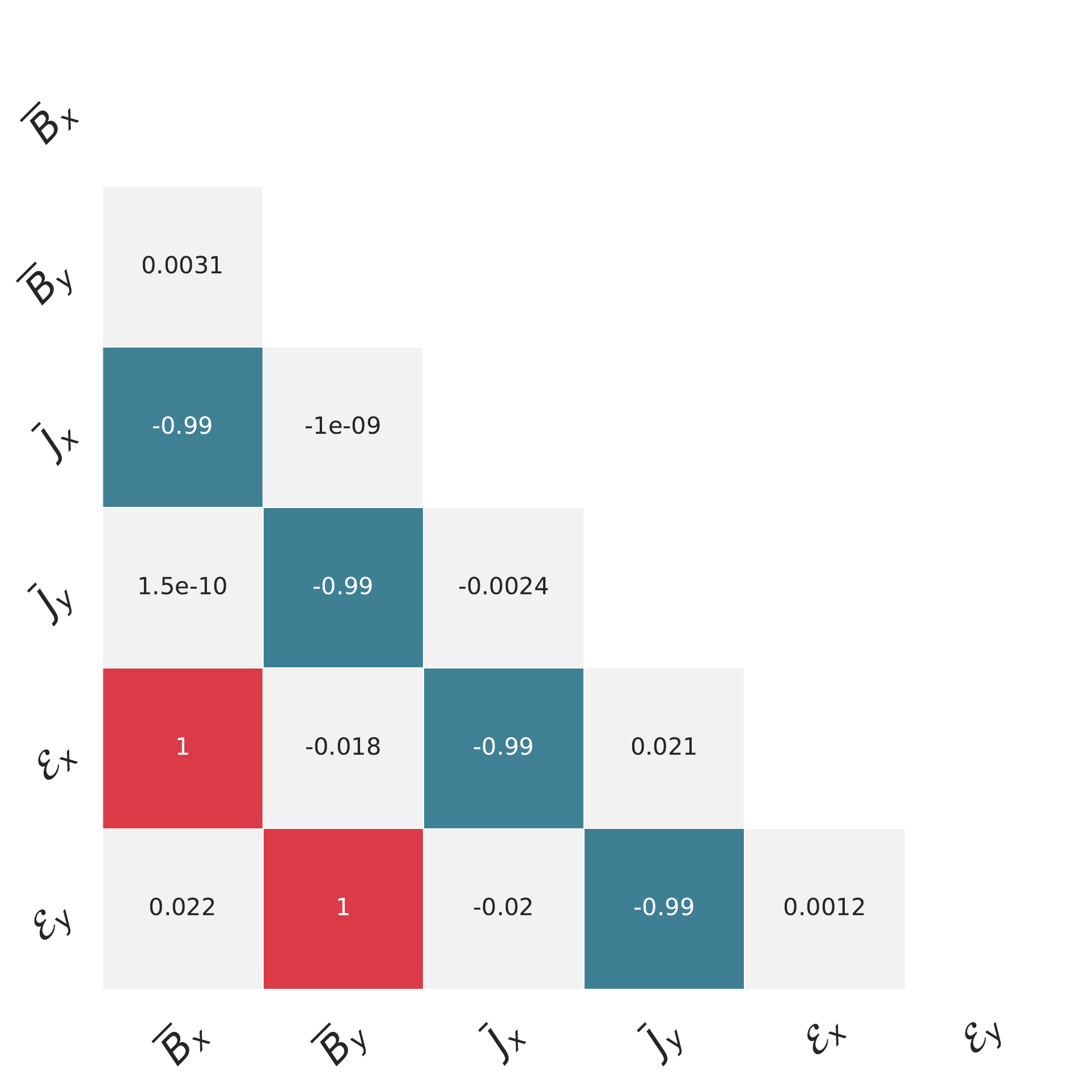}
	\caption{Heat map of correlation coefficients between the different fields: $Rm=10^3$ (left panel), $Rm=1.5\times 10^4$ (right panel). 
	This analysis implies that the $\emf_x$ has a strong correlation with $\Bbx, -\Jbx$ while $\emf_y$ has a strong correlation with $\Bby, -\Jby$. 
    Moreover, mean fields ($\Bbx, \Bby$) are not linearly independent of current densities ($\Jbx, \Jby$).
    }
	\label{fig:corr}
\end{figure*}


In Fig.~\ref{fig:corr}, we show the linear (Pearson) correlation coefficients among the various fields. The correlation between the current densities and the magnetic fields is nearly $1$ indicating they are linearly dependent. Strong linear correlation does not rule out the physical importance of a variable - it only implies that from a curve fitting point of view, it has redundant information. Since we are using helical forcing leading to a helical state, $\Jbar = \nabla \times \Bbar \sim \Bbar$. At high $Rm$, one might expect that the power in the velocity and magnetic field spectrum will be distributed across multiple modes implying weaker correlations. However, Fig.~\ref{fig:corr} (right panel) indicates the correlation between $\Bbar$ and $\Jbar$ is still perfect. This means that we can \textit{eliminate current density as an independent variable}. See, however, the work \cite{tilgner2008} that suggests that the tensorial form of $\alpha_{ij}, \eta_{ij}$ in the non-linear regime depends on $\mean{B}_i\mean{B}_j/\Bbar^2$. In keeping with previous work, for linear basis we do consider the current density as an independent variable but eliminate it for the non-linear (polynomial) basis where its inclusion will lead to several redundant terms.

We consider two sets of basis in this subsection: 
\begin{enumerate}
    \item Linear: $\Bbx$, $\Bby$, $\Jbx$, $\Jby$.
    \item Polynomial of $O(3)$: $\Bbx$, $\Bby$, $\Bbx \Bby$, $\Bbarsq$, $\Bbarsq \Bbx$, $\Bbarsq \Bby$.
\end{enumerate}
We note that for kinetically forced simulations, $v^2 \sim \text{constant}$, which means that the total energy conservation equation yields, $\Bbx^2 + \Bby^2 = \Bbarsq \sim \text{constant}$ in the saturated state. This makes $\Bbx^2$ and $\Bby^2$ linearly dependent terms and thus we do not consider them separately. Another motivation to use $\Bbarsq$ is because work on dynamo quenching often considers $\alpha \sim 1/(1+\Bbarsq/B_{eq}^2)$, which when expanded for the $\emfb$ leads to cubic terms of the form $\Bbarsq \mean{B}_i$ ($i=x,y$).
We will be using the publicly available \textsc{python} library \textsc{scikit-learn} (see \citealt{mueller, geron}, for introductions) for modeling the data.

Previous work on dynamo theory has often relied on linear regression methods \citep{bran2002,bran2018} with some effort to incorporate non-linear effects through magnetic helicity conservation (\citealt{pouquet1976,blackman2015}; see also non-linear test field method \citealt{rheinhardt2010}). In this section, we use linear regression with a regularization term for both the linear and polynomial basis discussed previously.

\begin{figure*}
	\centering
	\includegraphics[width=0.45\textwidth]{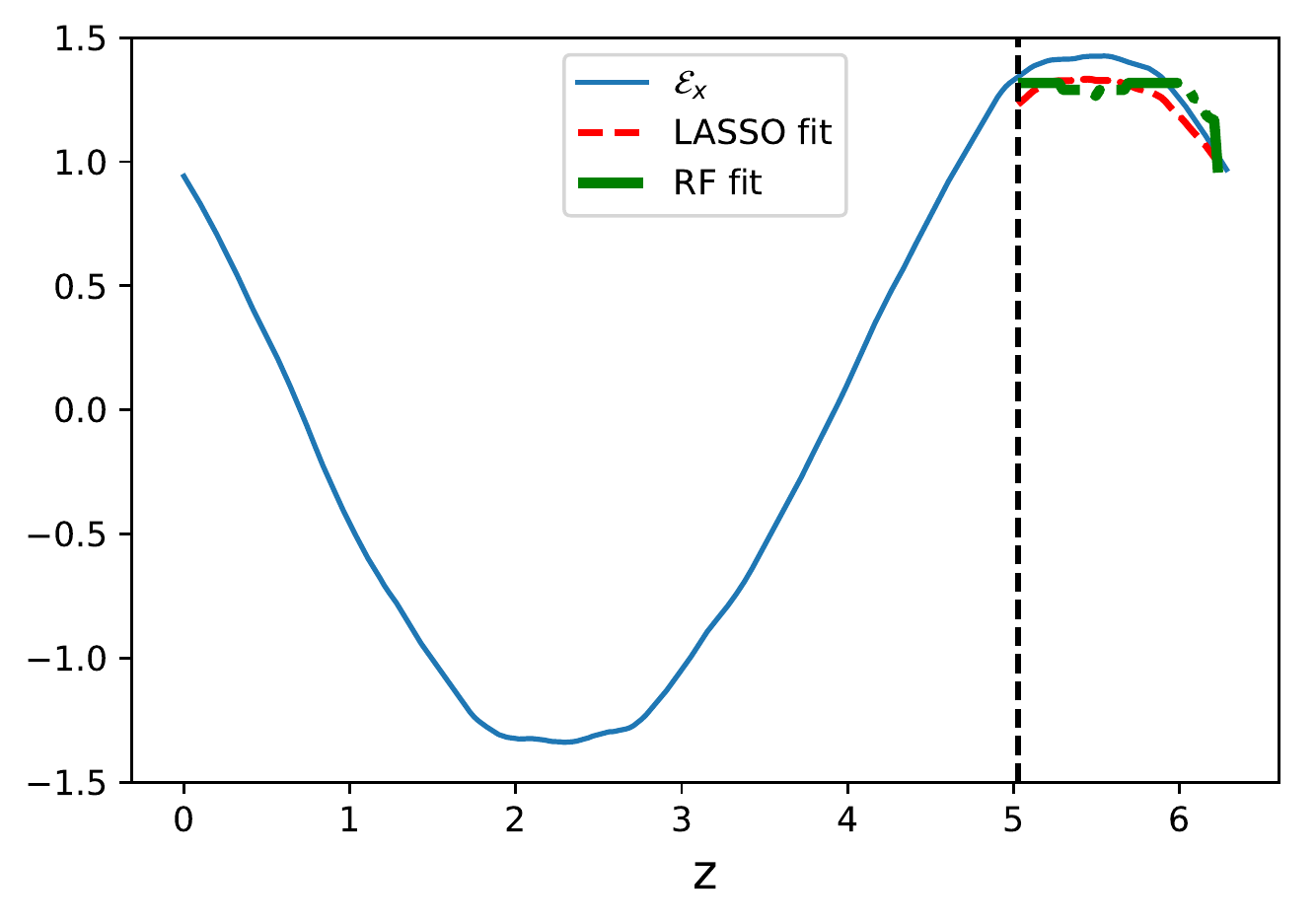}
	\includegraphics[width=0.45\textwidth]{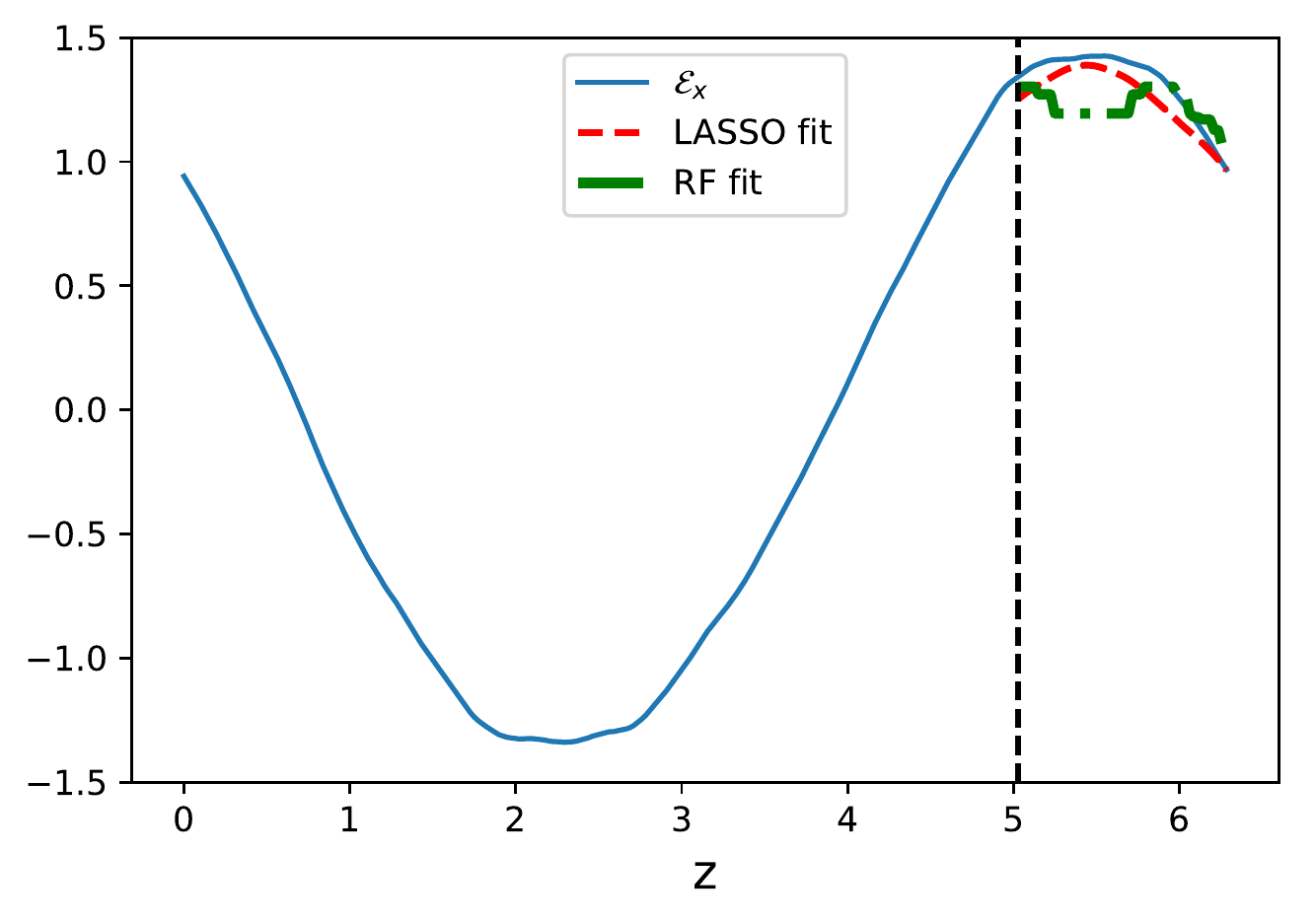}
	\caption{
		The fits for linear basis (left panel) and polynomial basis (right panel) for random forests and LASSO. Note that the horizontal lines seen in the fits are there because of decision tree regression: decision trees set one value of the $y$-axis for a given range of $x$-axis leading to a step-function like fit. In both cases, LASSO seems to do a better job at capturing the curved shape of the vertical profiles. The black vertical dashed line presents the train-test split ($80\%$ training data).
	}
	\label{fig:rfandlasso_space}
\end{figure*}

In Fig.~\ref{fig:rfandlasso_space}, we show the fits using random forests regression and LASSO. 
Because our data is small (only $256$ points), we use two-fold cross validation to determine various hyperparameters of LASSO and random forests. 
We found that bootstrapping leads to spurious results so we turned it off. 
A similar problem was found with random train/test splits: the spatial structure of the fields is like one full sinusoidal wave and randomly picking a fraction of the points leads to strange patterns. 
Moreover, while deep trees can lead to overfitting, in our case we found that a maximum depth of $8$ was optimal through 2-fold cross validation. 
The end results are still not as good as LASSO perhaps owing to the fact that the training set is only $204$ entries ($80\%$ of the data) - this is too low for an algorithm like random forest to really excel. 
That is the key lesson we have learnt while modeling this dataset throughout: ``small'' datasets mean simpler algorithms will outperform more sophisticated algorithms like random forests.


\begin{table*}
\begin{center}
    \begin{tabular}{c | ccccccccc} 
	Feature   &    OLS &  Lasso & Stability & RanFor & $RFE_{OLS}$ & $RFE_{rf}$ &    MIC &   Corr &   mean \\
    \hline
	$B^2$     & 0.0076 & 0.0048 &       0.0 & 0.0049 & 0.2 &    0.6     & 0.0000 & 0.1019 & 0.1149 \\
	$B^2 B_x$ & 0.6345 & 0.2577 &       0.0 & 1.0000 & 0.8 &    1.0     & 1.0000 & 0.9972 & 0.8362 \\
	$B^2 B_y$ & 0.0871 & 0.0240 &       0.0 & 0.0000 & 0.6 &    0.0     & 0.3979 & 0.1210 & 0.1538 \\
	$B_x$     & 1.0000 & 1.0000 &       1.0 & 0.4404 & 1.0 &    0.8     & 1.0000 & 0.9982 & 0.7793 \\
	$B_y$     & 0.0818 & 0.0000 &       0.0 & 0.0000 & 0.4 &    0.4     & 0.4048 & 0.1179 & 0.1756 \\
	$B_x B_y$ & 0.0000 & 0.0145 &       0.0 & 0.0001 & 0.0 &    0.2     & 0.2089 & 0.1366 & 0.0700 \\
\end{tabular}
\caption{
	Feature importances using various machine learning models and statistics.
	All the numbers are normalized such that they are positive and lie between $0$ and $1$ for the sake of comparison.
	Stability column refers to randomized LASSO where the LASSO shrinkage coefficient is randomly varied for different features and the feature that is most robust to this variation survives. 
	RFE stands for Recursive Feature Elimination where a model is trained with all features and the top ranking feature is given the most significance, while the bottom ranking one gets the smallest score. We applied this to both OLS and random forests, and reverse sorted the entries to give the highest score to top ranking feature.
	MIC stands for Maximal Information Coefficient and computes a normalized measure of the mutual information between two variables scaled between $0$ and $1$. It gives a quantitative measure of the question: how much information about some variable Y can be obtained through some variable X? 
	MIC is capable of capturing non-linear relationships that Pearson correlation (Corr) cannot. 
	In the last column we take the mean over all models that represents the synthesis of several models (linear, non-linear) to show that $\Bbx$ and $\Bbarsq \Bbx$ are the two strongest features.
	}
\label{table:feature_imp}
\end{center}
\end{table*}

In Table \ref{table:feature_imp}, we compare the coefficients from OLS, LASSO, randomized LASSO (stability), Recursive Feature Elimination (RFE) from OLS and $RFE_{rf}$ from random forests, Mutual Information Criterion (MIC), Pearson correlation (Corr) and the average of all these predictions in the last column. The meaning of these numbers is different. For OLS and LASSO, the numbers corresponding to the fields $\Bbx$ represent coefficients from linear regression and can be negative or positive. For random forests, the numbers represent the relative feature importance based on `mean decrease impurity' that is only positive. These feature importance numbers are scaled to sum to unity. \textit{Stability selection} is a general term that refers to checking robustness of model predictions by testing it on different random (bootstrapped) subsets of data with different hyperparameter values. In our particular application, we are using randomized LASSO that randomly varies the the LASSO coefficient \citep{meinshausen} \footnote{\url{https://blog.datadive.net/selecting-good-features-part-iv-stability-selection-rfe-and-everything-side-by-side/}; code: \url{https://github.com/scikit-learn-contrib/stability-selection}}.

RFE is a model-agnostic (wrapper) feature selection method that starts with a number of features and removes the most important features recursively. This is an example of ``backward'' feature selection \citep{guyon, hastie2008} and we used it with OLS and random forests in this case. Maximal Information Coefficient (MIC) is a measure of mutual information between two (random) variables and is capable of accounting for non-linear relationships that the Pearson correlation coefficient cannot (example: correlation between $x$ and $x^2$ will be nearly zero, but MIC will yield an answer close to one). We scaled the whole set of features $X_{\mathrm{train}}$ using standard scaling before calculating coefficients/feature importances. Moreover, because LASSO, OLS, Corr returned negative as well as positive values and RFE returned values ranging from $1$ to $6$, we used minmax scaling to re-scale all values between $0-1$. 

Ensembles of machine learning models can lead to improved accuracy \citep{wolpert}. Here we take the simplest possible approach: we take the unweighted average of various predictors to compute the overall score for each feature. While stability selection shows that $\Bbx$ is the dominant feature, the mean of all predictors gives $\Bbarsq \Bbx$ a slightly higher score. This could be because both $\Bbx$ and $\Bbarsq \Bbx$ have the same sinusoidal form. 

What do these results mean? Because we are dealing with low-dimensional data, constructing higher order terms from linear terms is not beneficial. LASSO is particularly well suited to identify important features in a low-dimensional model. It is, therefore, no surprise that the most important feature that (both randomized and ordinary) LASSO is picking up is the linear term $\Bbx$. Random forests, on the other hand, pick up $\Bbarsq \Bbx$ as the most significant feature and this might explain the poor prediction based on random forests seen in Fig.~\ref{fig:rfandlasso_space}. Random forests work best for large datasets; for small data that is really low-dimensional (that is nearly linear), random forests do not perform well. Therefore, the results of Table~\ref{table:feature_imp} are consistent with Fig.~\ref{fig:rfandlasso_space}. 

\subsection{Time series analysis: Vertically averaged data}
In contrast to the previous section, the magnetic fields and the EMF show exponential growth in time. To model this behavior, we found it convenient to do pre-processing where we took the logarithm of squared vertically averaged fields. Because of this choice, we do not consider polynomial expansions like we did before and instead focus on either the linear $\emfb \propto \Bbar$ form, or the quenched form $\emfb \propto \Bbar/(1 + \Bbarsq/B_{eq}^2)$.

\subsubsection{Preprocessing}

\begin{figure*}[t!]
	\centering
	\includegraphics[width=0.45\textwidth]{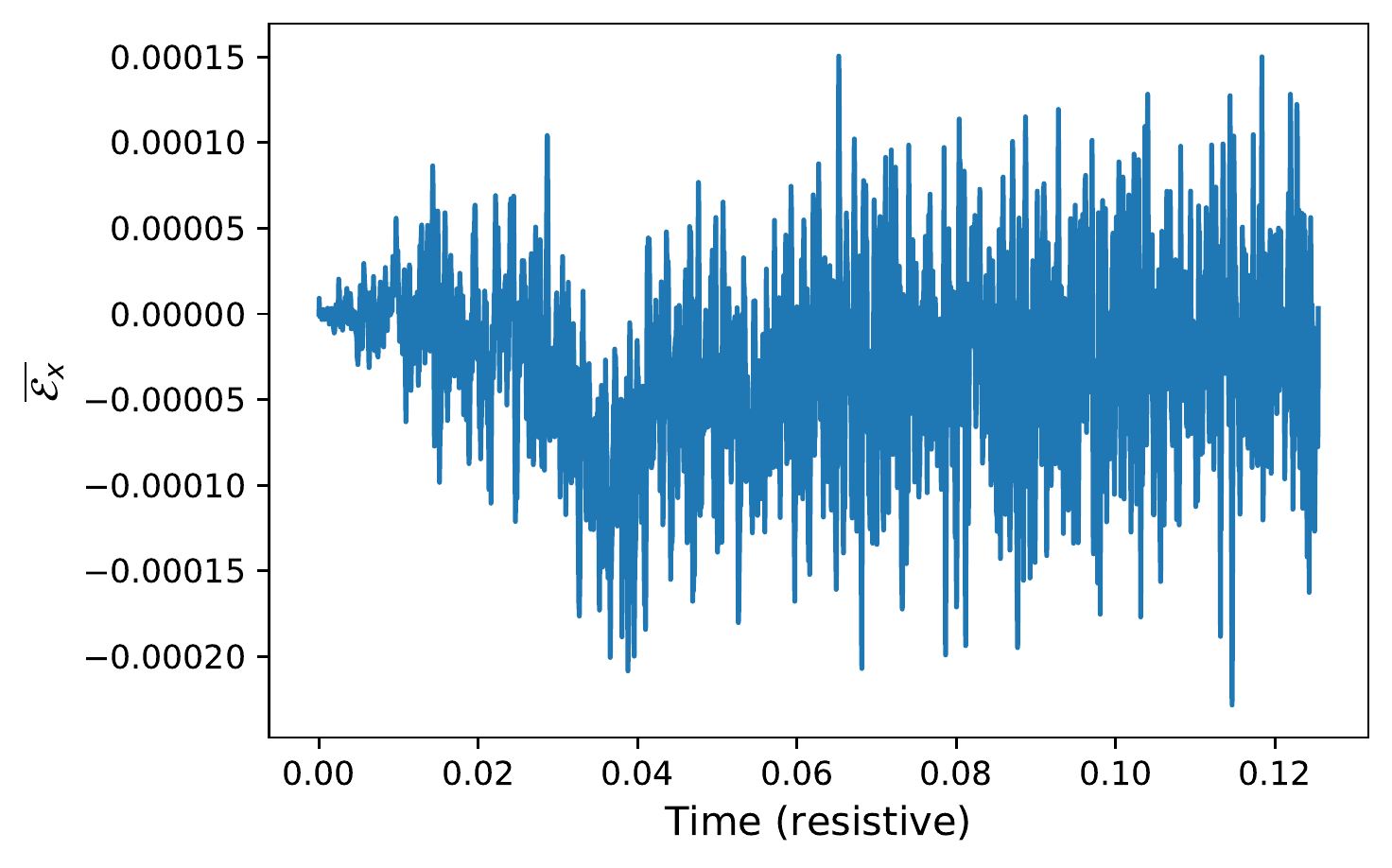}
	\includegraphics[width=0.45\textwidth]{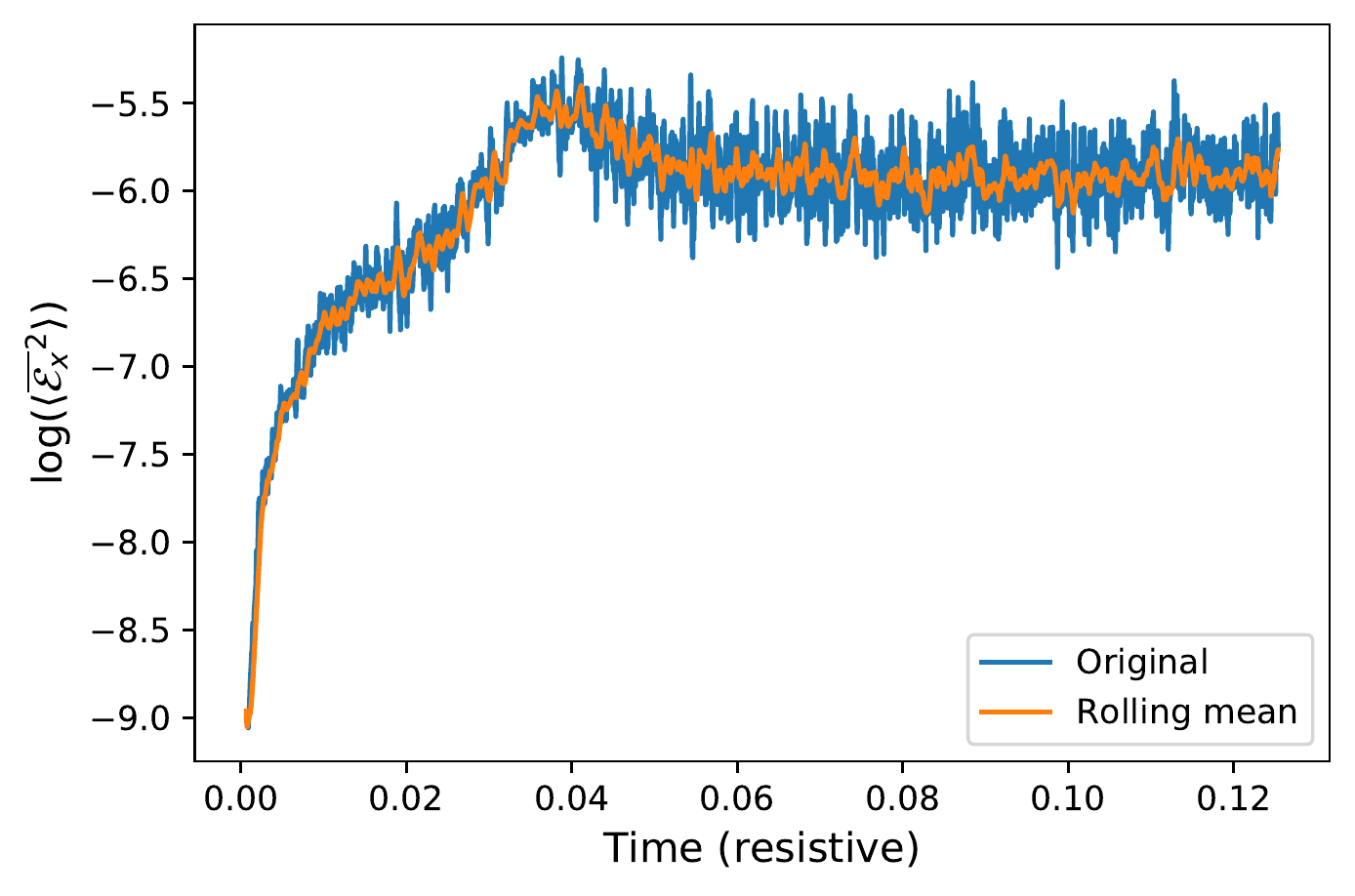}
	\caption{
        EMF evolution for $Rm = 1.5 \times 10^4$.
        Left: $\emf_x$ component is shown without any pre-processing ($z=\pi$). 
        Right: Here we show $\log\{ \langle\emf_x^2\rangle \}$, that is the squared EMF averaged over $z$. 
        Moreover, we use a moving average with window size $20$ for the orange line to smooth out the fluctuations.
	}
	\label{fig:emf_preprocess}
\end{figure*}

For the vertically averaged data set, we find that the fluctuations in the EMF are quite significant (see Fig.~\ref{fig:emf_preprocess}). Such large fluctuations make any kind of statistical analysis quite prohibitive. 
In this section, we therefore decided to use the log of $\emf_x^2$ averaged over the $z$ direction. 
Moreover, to reduce the fluctuations we used a moving window of size $20$ (right panel of Fig.~\ref{fig:emf_preprocess}). This allows for cleaner fits.

\subsubsection{LASSO and random forest}

\begin{figure*}[t!]
	\centering
	\includegraphics[width=0.45\textwidth]{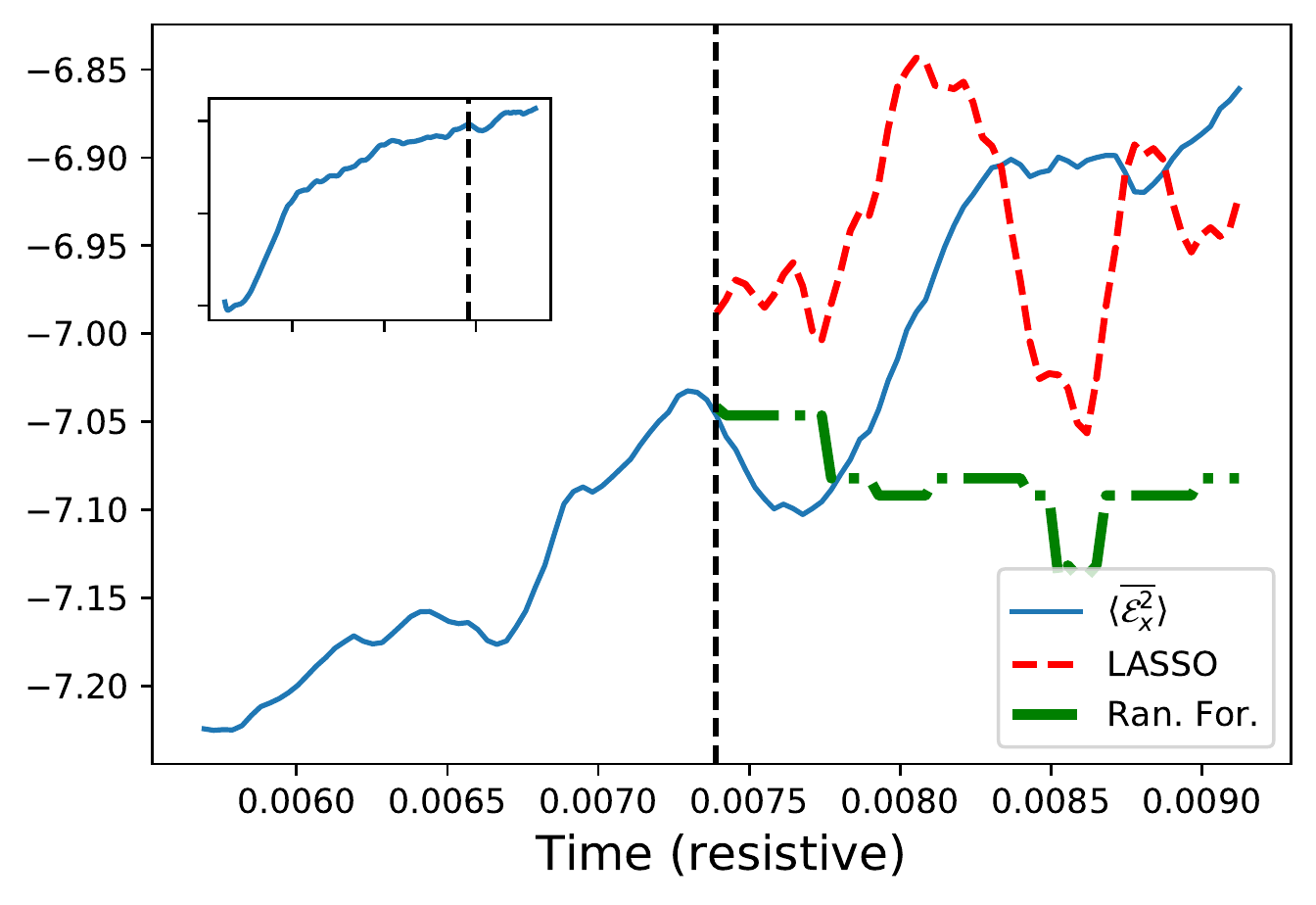}
	\includegraphics[width=0.45\textwidth]{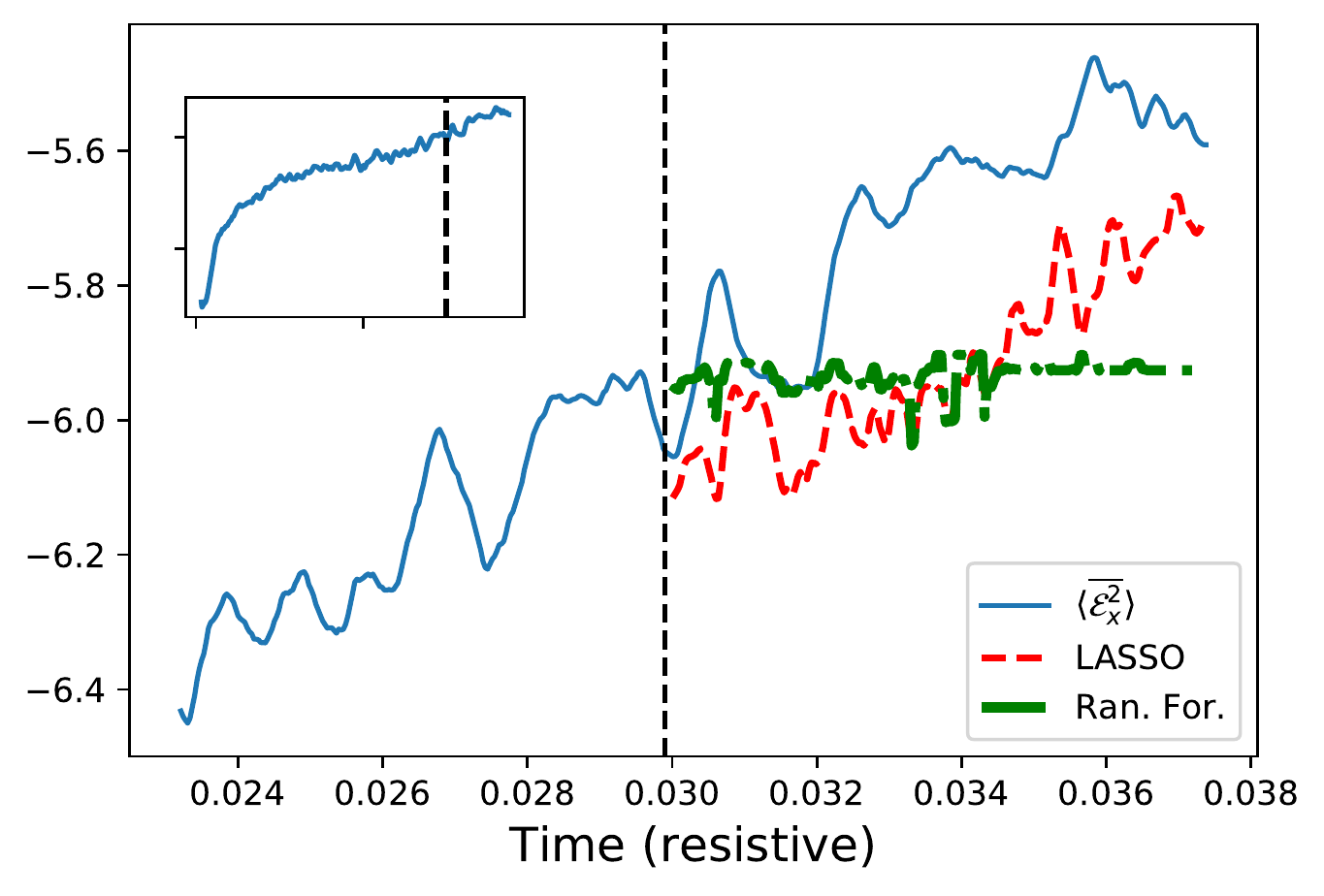}
	\caption{
		The fits for $\emf_x^2$ time series using LASSO and random forests for $Rm = 1.5\times10^4$. 
        Just like the fits for spatial data in Figs.~\ref{fig:rfandlasso_space}, we find that LASSO does considerably better than random forests, the latter of which again has a step like shape. \textit{\textbf{Left}}: we show the time series only up to the kinematic phase ($\sim 0.01$ resistive times) with the inset describing the train/test split. \textbf{\textit{Right}}: the time series is between kinematic and saturation phases. LASSO seems to capture the shape of the curve but is offset. In both cases, random forest prediction returns a nearly horizontal line, characteristic of decision tree regression. The black vertical dashed line presents the train-test split ($80\%$ training data).
	}
	\label{fig:rfandlasso_time}
\end{figure*}

In Fig.~\ref{fig:rfandlasso_time}, we show fits for the $\emf_x^2$ time series using LASSO and random forests. Time series data is particularly difficult to fit in the presence of fluctuations and trend as in the figure of $\emf_x^2$. In the kinematic phase (left panel), the random forest fit is not considerably different from LASSO fit whereas the random forest fit does poorly when the same time series is extended up to the saturation phase (right panel). 
This is a relatively common issue with random forests that they do not perform well when there is a ``trend'' present in the series. 
Just like the spatial profiles in last subsection, the LASSO outperforms random forests.

\subsubsection{Bayesian inference}

In this section we focus on reconstructing square of the ``true'' EMF field $\emfb^2$ as a function of time $t$.
As one of the simplest forms possible, we construct the reconstructed EMF, $\emfb_{\mathrm{R}}(z; t)^2 = \emf_{x,\mathrm{R}}(z; t)^2 + \emf_{y,\mathrm{R}}(z; t)^2$, from the corresponding magnetic field components as
\begin{align}
    \emf_{x, \mathrm{R}}(z; t) &= \frac{ \alpha }{1 + \Bbarsq(z; t)/B_{\mathrm{eq}}^2} \Bbx(z; t) \\ 
    \emf_{y, \mathrm{R}}(z; t) &= \frac{ \alpha }{1 + \Bbarsq(z; t)/B_{\mathrm{eq}}^2} \Bby(z; t),
\end{align}
where $\emf_{x, \mathrm{R}}^2$ and $\emf_{y, \mathrm{R}}^2$ are the reconstructed $x$ and $y$ components of the EMF field.
Here $\alpha$ and $B_{\mathrm{eq}}$ are our model parameters to optimize.
This EMF form has the nice property that in the limit $B_{\mathrm{eq}} \rightarrow \infty$ it reduces to the linear model.

After this, we are left with the freedom to define our likelihood or score-function that describes the quality of the reconstruction.
Here we define it via a simple residual function $\mathcal{R}$ 
\begin{equation}
\mathcal{R}= \frac{\emfb_R}{\emfb} + \frac{\emfb}{\emfb_R} -2,
\end{equation}
that penalizes the fit from over- and underestimating the reconstructed EMF.
Note that other forms of the residual functions are also possible.
For discrete simulation data results (defined on an array) the likelihood then simplifies to $\mathcal{L}_q = \sum_k |\mathcal{R}_k|$,
where the summation index $k$ is taken to be over the time dimension or the $z$ direction, depending on the type of fit we are considering.
Finally, when employing the MCMC fit note that we are actually minimizing $\log \sum_q \mathcal{L}_q$, as is typically done, to get a more shallow and slowly varying score-function.

\begin{figure*}
	\centering
    \includegraphics[width=0.75\textwidth]{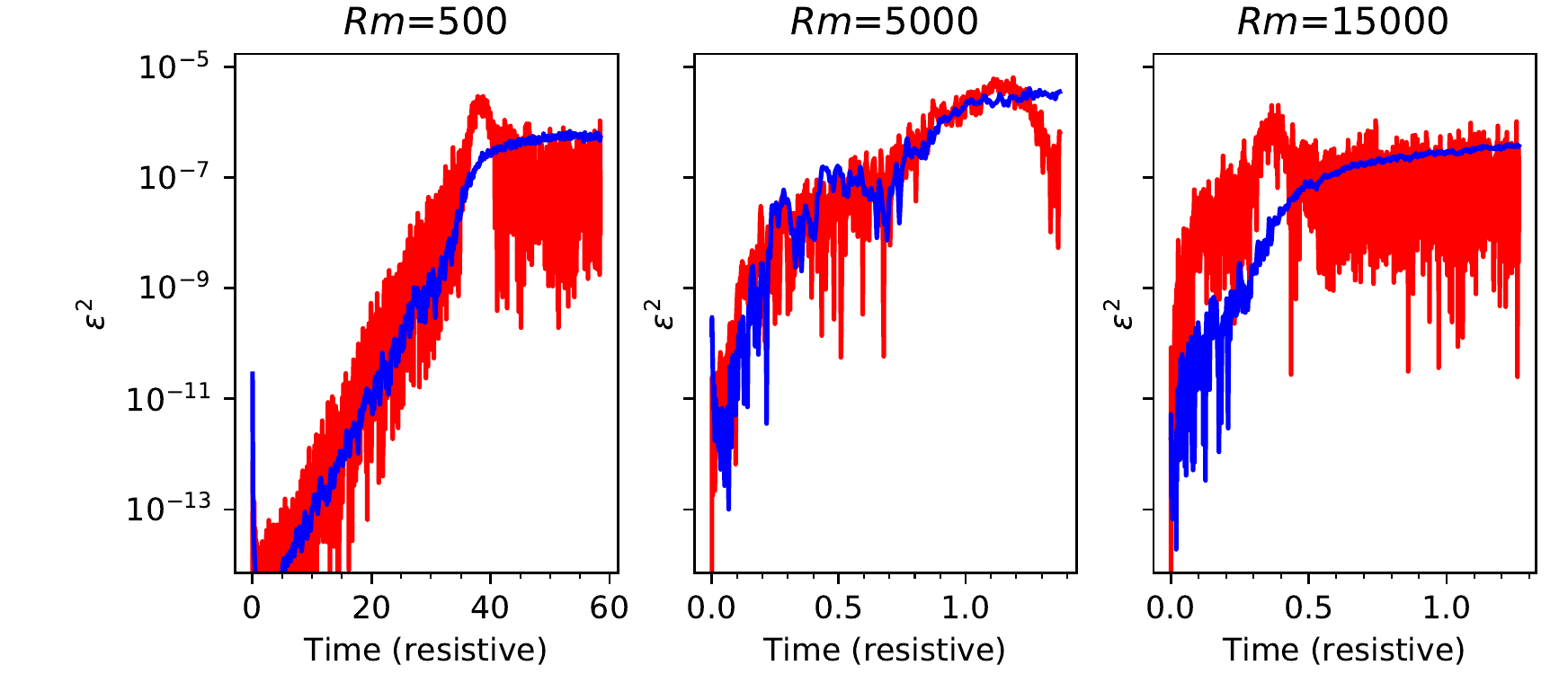}
	\caption{
        MCMC fits for the reconstructed $\emf^2$ fields.
        Three panel shows maximum likelihood Monte Carlo realizations of the fits (blue line) against the true $\emf^2$ curve (red line) for $Rm = 500$, $5 \times 10^3$, and $1.5 \times 10^4$.
    }
	\label{fig:mcmcfits}
\end{figure*}

Examples of the fit results are shown in Fig.~\ref{fig:mcmcfits} for three different $Rm$ values of $500$, $5\times10^3$, and $1.5\times10^4$ depicting the maximum a posteriori model.
The (quenched) EMF model does reproduce the qualitative behavior of the saturating $\Bbar$ field but especially for higher $Rm$ the fits become more worse.
Most importantly, we have found that constraints on the saturation field strength $B_{\mathrm{eq}}$ are always very loose.
Results without using the quenching form also lead to similar fits. 
This gives an indirect support for the previous machine learning results:
The data strongly favors isotropic $\alpha$ with a simplified model of $\emfb = \alpha \Bbar$.
The meaning of this `$\alpha$' could be confusing since we have already eliminated the mean current density $\Jbar$ as an independent variable. 
The $\alpha$ coefficient represents the saturated value where the growth term roughly balances the dissipation term, leading to the effective expression: $\alpha \sim k_1 (\eta + \eta_t)$. 

\subsection{Results summary}
For both the vertical and temporal profiles, various models suggest that $\emfb \propto \Bbar$ is a good description of the data. 
But if $d\Bbar/dt \propto \alpha \Bbar$, this will lead to exponential growth without any saturation. Is there an inconsistency here? In the induction equation, the fields at time `$t$' are related to the fields at earlier times. The machine learning models we use here are not capable of storing long term memory. The fits shown in this paper are comparing $\emfb$ and $\Bbar$ in some nearest neighbor sense and because both $\emfb$ and $\Bbar$ are growing exponentially before saturating at about the same time, they appear to be linearly correlated. Capturing the dynamical information requires more sophisticated algorithms that we leave for future work.

\section{Discussion}
\subsection{Caveats}

\begin{table*}
	\begin{center}
	\begin{tabular}{l|c|c|c|c}
		Property & OLS & LASSO & RF & Bayesian \\ 
		\hline 
		Non-linear & No & No & Yes & Yes \\ 
        Interpretable & Yes & Yes & Yes$^\mathrm{a}$  & Yes  \\ 
		Suitable for all data sizes? & Yes & Yes & Mid-sized to big data & Small to mid-sized data \\ 
        Interactions & No & No$^\mathrm{b}$ & Yes & Yes \\ 
		Prediction of trends & No & No & No & Yes \\ 
		Poor signal to noise ratio & No & No & Yes & Yes \\ 
	\end{tabular} 
	\caption{
        Summary of the advantages and disadvantage of different models considered in this work. 
        $^\mathrm{a}$Feature importances and partial dependence plots can help interpret random forests. 
        $^\mathrm{b}$Including cross terms like $\Bbx\Bby$ can help model interactions but it still assumes each feature is linearly independent.  }
	\label{table:models}
	\end{center}
\end{table*}

We summarize the properties of the different machine learning algorithms considered in this work in Table \ref{table:models}. Linear regression (both OLS and LASSO) are by definition capable of dealing with linear data only. However, they can handle some non-linear properties using a non-linear basis (for example, polynomial regression). The biggest advantage of linear models is the interpretability: the sensitivity of the target variable (EMF) with respect to the features ($\Bbx, \Bby$) is as easy as determining the coefficients of these terms. 
Linear models are also good at dealing with data of all sizes but matrix inverses are expensive for large data. Linear models might also be able to model trends in data if the features they consider have the same trend as the target variable. 

Random forests are based on ensembles of decision trees. While random forests are quite robust to outliers and missing data, the feature importances computed from random forests can be misleading \citep{strobl2008}. Moreover, random forests are not as interpretable as linear regression or single decision trees. A typical random forest fit can involve $10-100$ trees implying that interpretation is difficult. But computation of feature importances and partial dependence plots help in making sense of random forest results \citep{breiman2001two,hastie2008}. 
Random forests typically bootstrap data and use a random subset of features for each decision tree to determine which one leads to the lowest error. These two sources of randomness imply that random forests are best suited for intermediate to big data. Indeed in our ``small data'' case, we find that bootstrapping had to be turned off and the best fit models used less than $10$ decision trees.

Bayesian methods offer an interesting complementary toolkit to various machine learning techniques.
They also enable us to properly take the noisy nature of DNS into account. 
Here the big caveat is, that the model(s) needs to be known beforehand.
This, however, further arguments in favor of our hybrid approach applied in this paper.
We have used machine learning techniques to detect important features and then applied our physical knowledge to build valid models out from those.
Biggest caveat here is that Bayesian model optimization becomes more and more computationally demanding with multidimensional (and often multi-modal) data.
This puts a limitation on the real-life usability of the method when dealing with increasingly complex data.

The data considered in this paper is either spatially or temporally averaged reducing to $O(100-1000)$ entries. Big data typically refers to the case where the observations/rows are $\geq 10^6$. 
``Small'' data faces problems of overfitting and is more likely to be sensitive to outliers. Because of these issues, simpler models like linear regression with regularization tend to do well. It is, therefore, no surprise that LASSO has outperformed random forests. For larger datasets with irregular non-linear patterns, random forests and gradient boosting tend to do much better \citep{olson2017}. 

\subsection{Comparison with previous work}
We used the simplest model: isotropic with no non-local effects in either space and/or time. 
While earlier work has focused primarily on linear models of dynamos where $\emfb \sim \Bbar$ using linear regression and the test field method \citep{brandenburg2005}, non-linear models have also been considered \citep{rheinhardt2010}. 
We assumed isotropy, which could be a misleading assumption in the presence of shear (see \citealt{bran2002, karak2014}). 
Furthermore, in our model, we did not incorporate non-local effects that were considered by \citet{bran2002, hubbard2009, rheinhardt2012}. 
Combining machine learning algorithms with anisotropic, local, non-linear models might lead to new insights, and we leave this for future work.

The $\alpha^2$ dynamo considered in this work is an idealized system but has the advantage that due to maximally helical forcing and periodic boundary conditions, magnetic helicity is both conserved and is gauge invariant (see \citealt{blackman2015}). 
This simplistic setup allows developing analytical theory to explain the origin and saturation of magnetic fields \citep{pouquet1976,blackmanfield2002}. 
Numerical results seem to be consistent with theory \citep{brandenburg2001,brandenburg2012SSRv}. 
In this work, we considered existing analytical formulations and used machine learning tools to test whether the EMF is linear or non-linear in the mean magnetic fields.
Machine learning models can complement analytical theory in looking for reduced descriptions of high dimensional systems \citep{brunton_kutz_2019}.

\section{Conclusions}
Our main result can be summarized as follows: 
because we are in the ``small'' data regime and we are dealing with ordered data (high signal to noise ratio due to helically forced turbulence), regularized linear regression (LASSO) provides the best fit. For small organized data, sources of randomness such as random train/test splits, bootstrapping, cross validation with random subsets do not help. For this particular dataset, many of the features are strongly correlated with one another complicating the model fits even further. Strong correlations also imply that feature engineering to construct polynomial basis is not particularly useful. 

Random forests and MCMC do better with mid-sized to big data. Moreover, these ensemble methods tend to outperform linear and regularized linear regression when the signal to noise ratio is poor (see table 3). For the data we consider in this paper, it is therefore no surprise that LASSO provides the best fit.

What we intended to demonstrate using the example of $\alpha^2$ is that sophisticated machine learning algorithms can be used to analyze data from DNS of MHD turbulence thanks to the publicly available machine learning libraries. This provides an exciting opportunity to re-visit some long-standing problems in astrophysical flows where linear regression has dominated modeling efforts. For example, one can look at the full data cube without planar averaging and study whether 3D localized structures play a dominant role in the sustenance of the turbulence. 
Here, deep learning \citep{Lusch2018} will be a useful alternative as it is known to outperform classical machine learning methods in image processing, voice recognition, natural language processing. One particular problem of interest is shear-driven dynamos \citep{tobias2013, nauman1}.

\section*{Acknowledgments}
    We thank A. Brandenburg for several stimulating discussions and comments on an earlier draft of the paper.
    We also thank the anonymous referee for constructive comments that led to several improvements.
	The work has been performed under the Project HPC-EUROPA3 (INFRAIA-2016-1-730897), with the support of the EC Research Innovation Action under the H2020 Programme; in particular, the author gratefully acknowledges the support of Dhrubaditya Mitra (NORDITA, Stockholm) and the computer resources and technical support provided by PDC, Stockholm. We used the following python packages: \textsc{numpy} \citep{numpy}, \textsc{jupyter notebook} \citep{jupyter}, \textsc{matplotlib} \citep{matplotlib}, \textsc{pandas} \citep{pandas}, \textsc{scikit-learn} \citep{scikit-learn}, \textsc{emcee} \citep{emcee13}, \textsc{minepy} \citep{albanese}.


\label{lastpage}

\end{document}